\documentclass[aps,prd,twocolumn,superscriptaddress,groupedaddress, showkeys,floatfix]{revtex4} 
\usepackage{placeins}

\usepackage{subfig}
\usepackage{natbib}
\usepackage{mathtools}
\usepackage{scalerel}
\usepackage{graphicx}  
\usepackage{bm}        
\usepackage{tikz}
\usepackage{amssymb}   
\usepackage{lipsum}
\usepackage{caption}
\usepackage{lineno}
\usetikzlibrary{svg.path}
\usepackage{amsmath,array}
\usepackage{hyperref}
\hypersetup{
	colorlinks=true,
	linkcolor=blue,
	filecolor=magneta,      
	urlcolor=blue,
}
\DeclareCaptionJustification{justified}{\leftskip=0pt \rightskip=0pt \parfillskip=0pt plus 1fil}

\definecolor{orcidlogocol}{HTML}{A6CE39}
\tikzset{
    orcidlogo/.pic={
        \fill[orcidlogocol] svg{M256,128c0,70.7-57.3,128-128,128C57.3,256,0,198.7,0,128C0,57.3,57.3,0,128,0C198.7,0,256,57.3,256,128z};
        \fill[white] svg{M86.3,186.2H70.9V79.1h15.4v48.4V186.2z}
        svg{M108.9,79.1h41.6c39.6,0,57,28.3,57,53.6c0,27.5-21.5,53.6-56.8,53.6h-41.8V79.1z M124.3,172.4h24.5c34.9,0,42.9-26.5,42.9-39.7c0-21.5-13.7-39.7-43.7-39.7h-23.7V172.4z}
        svg{M88.7,56.8c0,5.5-4.5,10.1-10.1,10.1c-5.6,0-10.1-4.6-10.1-10.1c0-5.6,4.5-10.1,10.1-10.1C84.2,46.7,88.7,51.3,88.7,56.8z};
    }
}
\newcommand\orcidicon[1]{\href{https://orcid.org/#1}{\mbox{\scalerel*{
                \begin{tikzpicture}[yscale=-1,transform shape]
                \pic{orcidlogo};
                \end{tikzpicture}
            }{|}}}}

\begin{document}

\preprint{APS/123-QED}

\title{Revisiting constraints on superconducting cosmic strings in light of Dark Ages global 21-cm signal}
\author{Shibsankar Si$^{\orcidicon{0009-0001-8038-976X}}$\,}
\email{p21ph001@nitm.ac.in}
\affiliation{National Institute of Technology Meghalaya, Shillong, Meghalaya 793 003, India}

\author{Vivekanand Mohapatra$^{\orcidicon{0000-0002-5816-5225}}$\,}
\email{p22ph003@nitm.ac.in}
\affiliation{National Institute of Technology Meghalaya, Shillong, Meghalaya 793 003, India}

\author{Pravin Kumar Natwariya\,$^{\orcidicon{0000-0001-9072-8430}}$\,}
\email{pvn.sps@gmail.com}
\affiliation{School of Fundamental Physics and Mathematical Sciences, Hangzhou
Institute for Advanced Study, University of Chinese Academy of Sciences (HIAS-UCAS), Hangzhou, 310024, China}
\affiliation{University of Chinese Academy of Sciences, Beijing, 100190, China}
\affiliation{International Centre for Theoretical Physics Asia-Pacific (ICTP-AP), Beijing, 100190, China}

\author{Alekha C. Nayak$^{\orcidicon{0000-0001-6087-2490}}$\,}%
\email{alekhanayak@nitm.ac.in}
\affiliation{National Institute of Technology Meghalaya, Shillong, Meghalaya 793 003, India}%

\date{\today}

\begin{abstract}
The Superconducting Cosmic Strings (SCS) are a special case of cosmic strings that have a core carrying a charged field. When SCS passes through magnetized regions, the charged particles in the string experience a Lorentz force, which can produce radiation on the entire electromagnetic spectrum. This radiation can inject energy into the surrounding plasma, resulting in a modification of the thermal and ionization evolution of the intergalactic medium (IGM) and, subsequently, the global 21-cm signal. The signatures of SCS in the post-recombination era have been primarily studied in the low-frequency (radio) regime, which does not impact the state of the IGM. In this work, we study the effect of decaying SCS on the dark ages global 21-cm signal $(\delta T_b)$, considering both the ionizing and radio radiation. The dark ages signal can provide pristine cosmological information free from astrophysical uncertainties, as the universe was primarily homogeneous during this era in the absence of baryonic structure formation.  Considering a change in the $\delta T_b$ at redshift $z\sim 89$ from the $\Lambda \rm CDM$ framework, we derive an upper bound on the decay efficiency parameter, $g\equiv g(I,\,G\mu_s)$, to be $\lesssim 5.1\times10^{14}\, \rm GeV^2$, where, $I$ and $G\mu_s$ represent the loop current and string tension of SCS, respectively. 

 
\end{abstract}

\maketitle


\section{ Introduction}\label{sec:level1}
The absorption feature in the global 21-cm signal is one of the most exciting probes to constrain new physics after the recombination \cite{Pritchard:2011xb}. The 21-cm line originates from the hyperfine splitting in the 1S ground state of neutral hydrogen caused by the interaction of the proton and electron magnetic moments. The first-ever tentative detection of such an absorption signal in the 21-cm line was reported by the Experiment to Detect the Global Epoch of Reionization Signature (EDGES) in 2018 \cite{Bowman:2018yin}.  The observation suggests an absorption amplitude of $-0.5^{+0.2}_{-0.5}\, \rm K$ centered at redshift $z\sim 17$, which is larger by a factor of two than the one predicted by the studies based on $\Lambda\rm CDM$ framework of cosmology \cite{Furlanetto:2006jb, Pritchard:2011xb}. Previous studies have questioned the cosmological origin of this signal due to possible systematics and modeling assumptions \cite{Hills:2018vyr, Sims:2019kro, Singh:2019gsv, Bevins:2020jqf, Bradley:2018eev}. For instance, in the article \cite{Hills:2018vyr}, the authors showed that this signal can be reproduced by variation in the foreground modeling without requiring a true cosmological signal. In the article \cite{Sims:2019kro}, the authors investigate how systematics can produce absorption features mimicking a cosmological signal. In articles \cite{Singh:2019gsv, Bevins:2020jqf}, the maximum smooth function was used to show that the data may be compatible with foregrounds. In Ref. \cite{Bradley:2018eev}, the authors investigate whether the observed amplitude may result from artifacts within the ground plane. On the other hand, this anomalous detection suggests the existence of a colder intergalactic medium \cite{Barkana:2018qrx, Fialkov:2018xre, Slatyer:2018aqg, PhysRevD.98.103529}, or a hotter radio background \cite{Lawson:2012zu, Fraser:2018acy, Lawson:2018qkc, Pospelov:2018kdh, Fialkov:2019vnb, Li:2019loh, Acharya:2022vck, Caputo:2022keo, Dev:2023wel}. In the latter scenario, the radio radiation from the decaying superconducting cosmic strings (SCS) can explain the absorption amplitude reported by the EDGES \cite{Cyr:2023iwu}. Additionally, SCS can also explain the excess radio radiation observed by the Absolute Radiometer for Cosmology, Astrophysics, and Diffuse Emission 2 (ARCADE2) and Long Wavelength Array (LWA1) \cite{Fixsen:2009xn, Feng:2018rje, Dowell:2018mdb}. Similar to the EDGES observation, the ARCADE2 measurement is also debated in the literature due to uncertainties in galactic foreground modeling (see Ref. \cite{Subrahmanyan:2013eqa}). The LWA1 measurements have also been scrutinized in the literature due to their sensitivity to assumptions about Galactic emission modeling, as highlighted in Ref. \cite{Mondal:2020rce}. Recently, the Shaped Antenna Measurement of background Radio Spectrum-3 (SARAS-3) has rejected the existence of the entire signal with a 95.3\% confidence level after conducting an independent check \cite{Singh:2021mxo}. However, the presence of excess radiation in the early Universe cannot be completely ruled out. In this work, we discuss the effect of decaying SCS on the global 21-cm signal during the dark ages.

The theoretical and observational aspects of cosmic strings have been extensively studied in the literature \cite{Kibble:1976sj, Tashiro:2012nb, Vilenkin:1986zz, Hindmarsh:1994re, Vilenkin:2000jqa, Cyr:2023iwu, Cai:2011bi, Vachaspati:2008su, Berezinsky_2009, Babul:1987lza, Berezinsky:2001cp, Theriault:2021mrq, 10.1093/mnras/stae512, Hern_ndez_2014, Cyr:2023yvj}. In the article \cite{Kibble:1976sj}, the author was among the first to consider the existence of cosmic strings in the context of the spontaneous breaking of fundamental symmetry. As the universe evolves and cools down, cosmological phase transitions occur in the early universe. During the transitions, cosmic strings may form by spontaneously breaking fundamental symmetries in the early universe and may continue to exist today \cite{Tashiro:2012nb, Vilenkin:1986zz, Hindmarsh:1994re, Vilenkin:2000jqa}. The theoretical model of cosmic strings that interact gravitationally has been studied extensively in articles \cite{Vachaspati:1984gt, Hogan:1984is, Spirin:2020hca}. These types of string loops oscillate and decay by emitting gravitational radiation, which can be described by a single dimensionless parameter $G\mu_s$. Here, $G$ is Newton's gravitational constant and $\mu_s$ is the string tension. Additionally, several models have shown that some symmetry-breaking patterns can provide the superconducting properties to strings  \cite{Witten:1984eb, Vilenkin:1986zz, Ostriker:1986xc}. Usually, the string current $I$ and the dimensionless string tension $G\mu_s$ are the two parameters that characterize superconducting cosmic string (SCS) models. When SCS moves through a magnetized region, the charged particle present in the string experiences a Lorentz force, leading to the emission of electromagnetic radiation \cite{Vilenkin:1986zz, Berezinsky_2009}. The radiation from strings is not isotropic; however, it is most effective in cusps, substructures of strings that move almost the speed of light, and kinks, discontinuities in the string tangent vector \cite{Cyr:2023iwu}. When currents are present on strings, different observable signals can be investigated and probed by ongoing and future research \cite{Cyr:2023iwu, Cai:2011bi, Vachaspati:2008su, Berezinsky_2009, Babul:1987lza, Berezinsky:2001cp, Tashiro:2012nb, Theriault:2021mrq, 10.1093/mnras/stae512, Hern_ndez_2014, Cyr:2023yvj, Fixsen:2009xn}. Numerous observable signatures have led to draw constraints on the $G\mu_s$-$I$ parameter space. Cosmic strings can affect various cosmological phenomena. For instance, energy injection into plasma from SCS radiation can distort the cosmic microwave background (CMB) radiation \cite{Tashiro:2012nb, Cyr:2023iwu}, their potential role as gamma ray burst (GRB) that can be detectable by the existing instruments \cite{Babul:1987lza, Berezinsky:2001cp}, and radio emission from SCS could be a valuable probe to study for SCS and constrain various fundamental models \cite{Cai:2011bi, Vachaspati:2008su}, additionally it influence the production of extremely high energy neutrinos \cite{Berezinsky_2009}. The SCS can also affect the thermal and ionization evolution of the Universe during the post-recombination era.

During the post-recombination era,  especially during the dark ages and Cosmic Dawn era, SCS could produce both ionizing and nonthermal radio photons significantly \cite{Tashiro:2012nv}.  The effect of the excess radio photons from SCS on the global 21-cm signal has been studied in Refs. \cite{Theriault:2021mrq, 10.1093/mnras/stae512}. In the earlier article \cite{Hern_ndez_2014}, the global 21-cm signal was used to search for cosmic string wakes before the Epoch of Reionization. In the article \cite{10.1093/mnras/stae512}, the authors applied a Bayesian analysis considering the global 21-cm signal from SARAS-3, the upper bound on the 21-cm power spectrum from the Hydrogen Epoch of Reionization Array (HERA), and unresolved X-ray backgrounds from high-redshift galaxies to obtain an upper limit on the SCS parameter space. In Ref. \cite{Cyr:2023iwu}, authors investigated ``soft-photon heating" on the 21-cm signal due to excess radio radiation from SCS. Additionally, in the article \cite{Cyr:2023yvj}, they have also shown that considering the low-frequency spectrum from SCS can fit ARCADE2 measurement accurately \cite{Fixsen:2009xn, Feng:2018rje}. However, all of these studies have considered only the low-frequency regime of the entire electromagnetic spectrum produced from superconducting cosmic strings in drawing constraints on the loop current and string tension.

In this work, we consider the spectrum of photons in both the radio and the ionizing $(1\,\rm keV)$ regime. We then study the thermal and ionization evolution of the IGM and subsequent effects on the dark ages global 21-cm signal. The dark ages offer an astrophysically clean window to probe exotic physics. However, the detection of the corresponding low-frequency 21-cm signal is extremely challenging due to strong and poorly understood foregrounds, instrumental limitations, and particularly the Earth's ionosphere, which becomes opaque below $\sim40$ MHz \cite{Furlanetto:2006jb, Pritchard:2011xb}. Although future ground-based experiments, such as the proposed Mapper of the IGM Spin Temperature (MIST), aim to detect the signal in the $25–105$ MHz range \cite{Monsalve:2023lvo}, ionospheric distortion severely limits such efforts at frequencies below $\sim40$ MHz. Overcoming these challenges is one of the main motivations for several proposed lunar and space-based missions—such as FARSIDE \cite{farside}, DAPPER \cite{dapper}, FarView \cite{dapper}, SEAMS \cite{seams}, and LuSee-Night \cite{luseenight, 10906958}--- which are designed to operate above the Earth’s ionosphere and away from terrestrial radio frequency interference (RFI), thereby improving the chances of detecting the dark ages 21-cm signal. Lastly, we find that the cosmic strings that can explain the excess radio signal, shown in the previous studies \cite{Theriault:2021mrq, 10.1093/mnras/stae512, Cyr:2023iwu, Cyr:2023yvj}, can also produce enough ionizing radiation that can ionize the IGM during the dark ages and cosmic dawn. Therefore, we consider $5-15\,\rm mK$ change in the standard dark ages global 21-cm signal to draw an upper bound on the cosmic strings.

Throughout this paper, we work in the natural unit in which $c=\hbar=k_B=1$.  We use parameters in the context of standard $\Lambda\rm CDM$ cosmology: $\Omega_b=0.04859$, $\Omega_m=0.315$, $\Omega_r=10^{-4}$, $h=0.68$, and $H_0=100\times h$ \cite{Planck:2018vyg}. For the radiation-dominant era $(1+z)= \left({t'}/{t}\right)^{1/2}$, where $t'=1/(2H_0\sqrt{\Omega_r})$. In matter dominant era $(1+z)=\left({t_{eq}}/{t}\right)^{2/3}(1+z_{eq})$, where $z_{eq}=\left({\Omega_m}/{\Omega_r}\right)$.

This paper is organized as follows: we begin by discussing the energy deposition rate and the excess radio background from the SCS in Sec. \ref{sec.SCS}. Then, in sec. \ref{sec: 21-cm}, we briefly review the global 21-cm signal during the dark ages and the Cosmic Dawn era. In Sec. \ref{sec: gas evolution}, we briefly review the evolution of IGM temperature and ionization fraction in the presence of SCS decay. In sec. \ref{sec: results}, we have shown the IGM temperature evolution for different string loop currents and string tension values. Further, we evaluate the global 21-cm signal and constraints on the cosmic string parameter space from the 21-cm signal of the dark ages. Finally, in Sec. \ref{sec: conclusions}, we conclude our result with the existing constraints on SCS.

\section{Electromagnetic Radiation from superconducting cosmic string}\label{sec.SCS}

Cosmic strings are one-dimensional objects with a finite but very small width that may form during cosmological phase transitions. At any particular time, most of the string loops originate with approximately the same radius. At a given time, $t$, this radius is determined as a fraction of Hubble length, i.e. $L\approx \beta t$, where $\beta\approx O(10^{-1})$ \cite{10.1093/mnras/stae512}. After its formation, they oscillate with a time period $T \approx L$ and produce temporary substructures called kinks and cusps. The energy released by the string loop decay produces photons, gravitational waves, and exotic particles, resulting in the shortening of the loop size. All string loops emit gravitational radiation with the average power given by $P_g = \Gamma_g G \mu_s^2$ \cite{Tashiro:2012nb, Vachaspati:1984gt}, where $\Gamma_g \approx O(10^2)$ is the decay constant \cite{Vachaspati:1984gt}. Here, $G$ is Newton’s gravitational constant, and $\mu_s$ represents the string tension. In many different hypotheses beyond the Standard Model of particle physics, cosmic strings are considered to be superconducting \cite{Theriault:2021mrq, Witten:1984eb, Vilenkin:1986zz}. 
Electric currents are produced in the cosmic string when it passes through a magnetized region, releasing electromagnetic radiation. 
The average of the total power of electromagnetic radiation is given by $P_{em}=\Gamma_{em} I\sqrt{\mu_s}$ \cite{Tashiro:2012nb}. Here, $I$ is the current on the string, and $\Gamma_{em}\approx O(10)$ is a decay coefficient, which depends on the geometry of the loop \cite{Tashiro:2012nb, 10.1093/mnras/stad2457}.

For a given value of the string tension $ G \mu_s$, there exists a critical current $I^*$ at which gravitational radiation is equal to electromagnetic radiation. When the string current is greater than the critical current, electromagnetic radiation dominates over gravitational radiation. The critical current is given by \cite{Tashiro:2012nb}
\begin{eqnarray} \label{I*}
    I^* = \frac{\Gamma_g G \mu_s^{3/2}}{\Gamma_{em}}\, .
\end{eqnarray}

The SCS can decay via emitting gravitational and electromagnetic radiation. Therefore, the overall dimensionless decay rate is given by \cite{Cyr:2023iwu}

\begin{eqnarray} \label{decay_rate}
    \Gamma G \mu_s = \frac{(P_g + P_{em})}{\mu_s}=\Gamma_gG\mu_s+\frac{\Gamma_{em}I}{\sqrt{\mu_s}}\, ,
\end{eqnarray}
where, the dimensionless decay coefficient $\Gamma$ is a function of the string tension and the current on the string. The decay coefficient $\Gamma$ can be expressed as
\begin{eqnarray} \label{Gamma1}
 \Gamma=  \Gamma_g \left( 1+ \frac{I}{I^*}\right)\, ,
\end{eqnarray}
 where,
 
\begin{eqnarray} \label{Gamma2}
       \Gamma= 
\begin{dcases}
    \Gamma_g&     \text{for }I\ll I^*,\\
    \Gamma_g \left(\frac{I}{I^*}\right)    & \text{for }I \gg I^*\, .
\end{dcases}
\end{eqnarray}
If the initial length of a loop is $L_0$, then the length of the loop varies with time in the following ways \cite{Tashiro:2012nb, Cyr:2023iwu}

\begin{eqnarray} \label{L}
    L = L_0 - \Gamma~G \mu_s(t-t_0)
\end{eqnarray}

where $t_0$ is the initial time. If we assume $t\gg t_0$, and a slow decay rate, then $L_0$ can be expressed as \cite{Tashiro:2012nb}
\begin{eqnarray} \label{L_0}
    L_0=L+\Gamma~G \mu_s~t\, .
\end{eqnarray}
The differential number density of cosmic strings with initial loop length $L_0$ in the radiation-dominated epoch ($t\leq t_{eq}$) and matter-dominated epoch ($t>t_{eq}$) is given by \cite{Tashiro:2012nb}

\begin{eqnarray} \label{dN}
       dN= 
\begin{dcases}
    \frac{\kappa}{t^{(3/2)} L_0^{5/2}} dL &  \text{for }t\leq t_{eq}\, ,\\
    \frac{\kappa \beta}{t^{2} L_0^2}  dL            & \text{for } t> t_{eq}\, .
\end{dcases}
\end{eqnarray}
 where $\kappa\sim 1$ is a dimensionless electromagnetic emissivity constant, and $\beta = 1+\sqrt{t_{eq}/{L_0}}$ is also a dimensionless model-specific parameter, where the second term of $\beta$ accounts for the string loops in the matter-dominated era that are left over from the radiation-dominated era \cite{Tashiro:2012nv}.

Superconducting cosmic strings with cusps can emit an entire spectrum of electromagnetic radiation. Therefore, the energy injected into the primordial plasma and CMB can be thermalized efficiently before the recombination epoch $(z\sim 1100)$, consequently distorting the CMB spectrum. However, after the recombination epoch, the ionization fraction $(x_e)$ falls drastically, reaching $x_e\sim 10^{-3}$ at $z\sim 600$. Therefore, in the pre-reionization era $(10\lesssim z\lesssim 1100)$, cosmic strings can inject ionizing photons into the IGM, which could modify the thermal history, especially during the dark ages. The spectrum (number of photons per unit time per unit frequency) of photons with frequency $(\omega)$ emitted from a superconducting cosmic string with cusps can be expressed as \cite{Tashiro:2012nv},

\begin{equation}
    \dot{N}_{\omega}\equiv \frac{d^2N}{d\omega\,dt} \sim \frac{4\pi}{3}\,\frac{I^2L^{1/3}}{\omega^{5/3}}.
\end{equation}
The spectrum of photons emitted per unit volume in the matter-dominant era $(z<z_{\rm eq})$ is expressed as \cite{Tashiro:2012nv},

\begin{eqnarray}
    \dot{\mathcal{N}}_{\omega}  & = &\int_{0}^{\infty} dN (L,t)\,\dot{N}_{\omega},\nonumber\\
    &\approx & \frac{8\pi C}{9}\left(\frac{t_{\rm eq}}{t}\right)^{1/2}\frac{I^2}{(\Gamma~\mu_s\,\rm G)^{7/6}t^{8/3}}\,\omega^{-5/3},
    \label{eq: cs_photon_spectrum}
\end{eqnarray}
where, $C$ is the average number cusps in a loop, and $z_{\rm eq}(t_{\rm eq})$ is the redshift (time) at which $\Omega_r = \Omega_m$. The $\dot{\mathcal{N}}_{\omega}$ depends upon the frequency as $\omega^{-5/3}$. Therefore, the emitted spectrum of photons falls greatly in the higher frequency range (Eq. \ref{eq: cs_photon_spectrum}). To calculate the total energy density rate, we integrate Eq. \eqref{eq: cs_photon_spectrum} with frequency $(\omega)$ between $0$ to $\omega$. Thus, the volumetric energy rate injection can be expressed as,  

\begin{eqnarray}
     \frac{d^2E}{dVdt} & \equiv &\int_{0}^{\omega} \omega\dot{\mathcal{N}}_{\omega}\,d\omega \nonumber\\
    &=& \frac{8\pi C}{3}\left(\frac{t_{\rm eq}}{t}\right)^{1/2}\frac{I^2}{(\Gamma~\mu_s\,\rm G)^{7/6}t^{8/3}}\,\omega^{1/3}\, .
    \label{eq: volumeteric_energy_injection}
\end{eqnarray}
 Eq.~\eqref{eq: volumeteric_energy_injection} clearly shows a degeneracy between the string current $I$ and tension $G\mu_s$. To make this dependence explicit, we introduce a new decay efficiency parameter $g(I, G\mu_s)$ defined as
\begin{eqnarray}
g(I, G\mu_s) \equiv I^2 \big(\Gamma~G\mu_s\big)^{-7/6}, 
\label{eq:degeneracy_para}
\end{eqnarray}
And parameterize the time dependence part as
\begin{eqnarray}
A(t) = \frac{8\pi C}{3}\left(\frac{t_{\rm eq}}{t}\right)^{1/2}t^{-8/3} .
\label{eq:At}
\end{eqnarray}
In terms of these parameters, the energy injection rate in Eq. (10) can be written as
\begin{eqnarray}
\frac{d^2E}{dVdt}= A(t)~g(I, G\mu_s)~\omega^{1/3}, 
\label{eq: energy_injection_param}
\end{eqnarray}
In this work, we fix $C = 1$, representing the existence of at least one cusp per loop. Furthermore, $t_{\rm eq}/t$ can be expressed as $\left[(1+z)/(1+z_{\rm eq})\right]^{3/2}$, where $z_{\rm eq} = \Omega_{m0}/\Omega_{r0}$, and $t = (2/3)\,(\sqrt{\Omega_m}\rm H)^{-1}$ in a matter-dominated universe.

For a constant energy injection, the decay efficiency parameter $g(I, G\mu_s)$ becomes constant, assumed to be $g=g_0$. The resulting degeneracy then appears as a straight line on a log-log plot of the $I$ and $G\mu_s$. The slope of the straight line varies for two different regimes, $I<I^*$ and $I>I^*$, due to the different functional form of $\Gamma$ in Eq. \eqref{Gamma2}. The transitions between these two regimes occur along the curve defined by $I=I^*(G\mu_s)$ (shown as the orange dashed line in Fig.~\ref{plot: contour}). For the regime $I<I^*$, we obtain  
\begin{eqnarray} \label{eq: gmu_I<I*}
    G\mu_s=\frac{g_0^{-6/7}}{\Gamma_g}I^{12/7}
\end{eqnarray}
which corresponds to a straight line with slope $12/7$ (see Fig.~\ref{plot: contour}).
For $I>I^*$, the expression becomes
\begin{eqnarray} \label{eq: gmu_I>I*}
    g_0= \mathrm{constant}~\times~ I^{5/6} (G\mu_s)^{7/12}
\end{eqnarray}
where the constant depend on $G$ and $\Gamma_{em}$. In this regime, we obtain
\begin{eqnarray}
    G\mu_s\propto I^{-10/7}, 
\end{eqnarray}
so that the corresponding slope is $-10/7$ (see Fig.~\ref{plot: contour}).

From Eqs. \eqref{dN} and \eqref{eq: volumeteric_energy_injection}, we can conclude that, even though $d^2E/dVdt\propto \omega^{1/3}$ but the number density falls as $\omega^{-5/3}$ for higher frequencies, whereas both scales as $(1+z)^{19/3}$. Therefore, we might find a larger number of low-frequency photons than the high-frequency ones in the early universe. However, the existence of ionizing photons emitted by the loops cannot be ruled out \cite{Tashiro:2012nv}. 

We calculate the energy injection rate per unit volume $(d^2E/dVdt)$ in the entire frequency spectrum. We note that the entire spectrum of photons cannot ionize or heat the IGM. The authors in article \cite{Venumadhav:2018uwn} have shown IGM heating by Lyman alpha photons, making radio photons as a conduit--- which was later challenged in article \cite{Meiksin:2021cuh}. Additionally, in article \cite{Cyr:2023iwu}, authors have shown heating of IGM by soft photons via the free-free process, which requires detection of 21-cm signal and/or CMB spectral distortion in frequency $\nu<60\,\rm GHz$ \cite{Acharya:2023ygd}. In the present work, we have not considered either of these cases. Consequently, we segmented the energy density rate in Eq. \eqref{eq: volumeteric_energy_injection} into radio, non-ionizing, and ionizing/heating photons, rewriting as,

\begin{alignat}{2}
    \frac{d^2E}{dVdt}&   = A(t)~g(I, G\mu_s)  \bigg[\underbrace{\omega_{21 \rm{cm}/5.87\,\mathrm{\mu eV}}^{1/3}}_{\text{radio}} \nonumber\\
     &+\underbrace{ \left(\omega_{13.6\,\mathrm{eV}}^{1/3} - \omega_{21 \rm{cm}/5.87\,\mathrm{\mu eV}}^{1/3}\right)}_{\text{non-ionizing}} +  \underbrace{\left(\omega_{10^4\,\mathrm{eV}}^{1/3} - \omega_{13.6\,\mathrm{eV}}^{1/3}\right)}_{\text{ionizing/heating}}\bigg]\, .
    \label{eq: cs_tot_energy}
\end{alignat}
Here, the bracketed terms on the RHS of the equation represent angular frequencies (or energies) of different photon spectra. The first term $(\omega_{21\rm cm/5.87\,\rm \mu eV})$ represents the angular frequencies of photons with wavelength 21-cm. $\omega_{13.6\,\rm eV}$ and $\omega_{10^4\,\rm eV}$ represent the angular frequencies of photons with energy $13.6\,\rm eV$ and $10^4\,\rm eV$, respectively. Nonthermal radio photons with a wavelength of 21-cm can increase the background radiation temperature. In contrast, photons with energy $<13.6\,\rm eV$ cannot ionize hydrogen atoms in the ground state \cite{PhysRevD.70.043502}. However, soft X-ray photons with energy $\sim 10^4\,\rm eV$ have a large photoionization cross section and can ionize hydrogen atoms \cite{PhysRevD.70.043502}.

Now, to evaluate the deposition of energy to heat and ionize the IGM, we consider the last term of the above equation, that is

\begin{alignat}{2} 
    \frac{d^2E}{dVdt}\bigg{|}_{\text{dep}} = \mathcal{F}(\omega, z)\, A(t)~g(I, G\mu_s)
     \left\{\omega_{10^4\,\mathrm{eV}}^{1/3} - \omega_{{13.6}\,\mathrm{eV}}^{1/3}\right\},
    \label{eq: energy deposition}
\end{alignat}
where $\mathcal{F}(\omega, z)$ represents the fraction of energy deposition with respect to the injected energy \cite{PhysRevD.87.123513, PhysRevD.93.023527, PhysRevD.93.023521}. The energy density of radio photons at time $t$ resulting from decaying SCS was thoroughly analyzed in Ref. \cite{Theriault:2021mrq, 10.1093/mnras/stae512}. It is calculated by integrating the volumetric energy injection rate over a time $t$ and can be expressed as \cite{10.1093/mnras/stae512},

\begin{eqnarray}\label{rho21}
    \rho_{21}(t)&=&\int dt\, \frac{d^2E}{dVdt}\bigg{|}_{\omega_{5.87 \rm{\mu eV}}}  \nonumber\\
    &=& \int A(t)~g(I, G\mu_s)\,\omega_{5.87\,\mathrm{\mu eV}}^{1/3}\, dt\, .
\end{eqnarray}
The excess radio background generated from decaying SCS at the 21-cm line frequency $\omega_{21}$ can be defined as
\begin{eqnarray}\label{t21}
    T_{21}^{\rm{SCS}}=\frac{3\pi^2}{\omega_{21}^3}\rho_{21}(t)\, .
\end{eqnarray}
The effective background photon temperature at the 21-cm wavelength can now be expressed as $T_R=T_{21}^{SCS}+T_\gamma$, where $T_\gamma$ is the CMB temperature. In the further sections, we formulate the impact of energy injections on the IGM temperature and global 21-cm signal.

\section{Global 21-cm absorption signal} \label{sec: 21-cm}

The baryon content of the universe in the post-recombination era primarily consisted of neutral hydrogen and a fraction of helium atoms. Due to the spin interaction between the electron and proton, the ground state of the neutral hydrogen atom splits into two states--- the singlet $(\rm F= 0)$ and triplet $(\rm F = 1)$ states. The relative population density of the singlet $(n_0)$ and triplet $(n_1)$ state is defined as ${n_1}/{n_0} = {g_1}/{g_0}\,\exp[-T_*/T_s]$, where $g_1= 3$ and $g_0 =1$ are the statistical weights of the respective states. $T_* = 68\, \rm mK$ is the equivalent temperature of the photons produced from the transition between the singlet and triplet states. These photons have a frequency of 1420 MHz or a wavelength of 21 cm. $T_s$ represents the spin temperature that determines the relative population density \cite{4065250, Pritchard:2011xb}. 

The redshifted difference between the $T_s$ and radio background temperature $(T_{R})$ is defined as the brightness temperature $(\delta T_b)$, which is expressed as $\delta T_b = [(T_s-T_{R})/1+z]\, (1-e^{-\tau_{21}})$. Here $\tau_{21}$ is the optical depth of 21 cm photons. In the limit $\tau_{21}\ll 1$, the brightness temperature or the global 21-cm signal can be expressed as \cite{ DAmico:2018sxd, Mitridate:2018iag, Cyr:2023iwu, PhysRevD.110.123506, Nishizawa:2024bnh}

\begin{equation}\label{T21}
    \delta T_b \approx 27x_{\rm HI}\left(1-\frac{T_R}{T_s}\right)\left(\frac{0.15}{\Omega_m}\frac{1+z}{10}\right)^{0.5}\left(\frac{\Omega_bh}{0.023}\right)\mathrm{mK}\, .
\end{equation}
Where the neutral hydrogen fraction $x_{\rm HI}=\frac{n_{\rm HI}}{n_H}$, $n_H$ is the total hydrogen number density, and $n_{\rm HI}$ is the neutral hydrogen number density. The evolution of $T_s$ is given by \cite{Venumadhav_2018, Furlanetto:2006tf, Acharya:2022txp}
\begin{equation}
    T_s^{-1} = \frac{T_R^{-1}+x_c{T_{gas}}^{-1}+x_\alpha T_\alpha^{-1}}{1+x_c+x_\alpha}\, ,
\end{equation}
where  $T_{gas}$ and $T_\alpha$ are the IGM and the color temperature, respectively. Here, $x_c$ and $x_\alpha$ are the collisional and Wouthuysen-Field coupling coefficients, respectively \cite{1952AJ.....57R..31W, 1959ApJ...129..536F, 1958PIRE...46..240F}. From Eq. \eqref{T21}, we can observe that one expects an absorption signal for $T_s< T_R$. Below, we will explain the two absorption troughs expected in the $\Lambda\rm CDM$ framework.

\subsection{Dark Ages signal}

After the epoch of recombination $(z\approx 1100)$, the IGM was coupled to the CMB via Inverse Compton scattering between the electrons/protons and the CMB photons. Therefore, the IGM and the CMB shared the same temperature resulting in $\delta T_b = 0$. After $z\sim 200$, the inverse Compton scattering becomes ineffective, and the IGM and CMB temperatures evolve as $(1+z)^2$ and $(1+z)$, respectively, due to the universe's adiabatic expansion. The collisional coupling $(x_c)$ between neutral hydrogen atoms and electrons/protons was efficient, which kept $T_s$ coupled to the IGM temperature till $z\sim 40$. Therefore, we expect an absorption trough at redshifts $z\sim 200-40$--- termed as the dark ages global 21-cm signal. The collisional coupling is defined as \cite{ 2001A&A...371..698L, 2006ApJ...637L...1K, Furlanetto:2006tf, Pritchard:2011xb},
\begin{equation*}
    x_c = \frac{T_*}{T_R}\,\frac{n_ik^{i\rm H}_{10}}{A_{10}},
\end{equation*}
where $n_i$ represents the number density of the
species ``$i$" present in the IGM while $k^{i\rm H}_{10}$ represents their corresponding collisional spin deexcitation rate. $A_{10} = 2.85 \times 10^{-15}\,\rm Hz$ is the Einstein coefficient for spontaneous
emission in the hyperfine state. The deexcitation rates $k^{HH}_{10}$ and $k^{eH}_{10}$ can be approximated in a functional form as follows \cite{ 2001A&A...371..698L, 2006ApJ...637L...1K, Pritchard:2011xb, PhysRevD.110.123506}

\begin{alignat}{2}
	k^{HH}_{10} & = 3.1 \times 10^{-17}\left(\frac{T_{gas}}{\mathrm{K}}\right)^{0.357}\cdot e^{-32\mathrm{K} / {T_{gas}}}, \\
	\log_{10}{k^{eH}_{10}} & = -15.607 + \frac{1}{2}\log_{10}\left(\frac{{T_{gas}}}{\mathrm{K}}\right)\times \nonumber \\ & ~~~~\qquad\qquad \exp\left\{-\dfrac{\left[\log_{10} \left({T_{gas}}/\mathrm{K}\right)\right]^{4.5}}{1800}\right\}\, .
\end{alignat}
All $k^{iH}_{10}$ terms have the dimension of $\rm m^3 \rm s^{-1}$. Here, $k^{iH}_{10}\rm s$ have been approximated under the consideration that $T_{gas}<10^4\, \rm K$. Further, at redshifts $z\lesssim 40$, $T_s$ approaches the CMB temperature as $x_c$ become $\ll 1$, which led to $\delta T_b\sim 0$. However, the formation of astrophysical structures in the early universe can emit Lyman alpha $(\rm Ly\alpha)$ radiation that can alter $T_s$ in the later time. Below, we discuss the effect of star formation on $T_s$ and thereby on $\delta T_b$. 

\subsection{Cosmic dawn signal}\label{sec: cosmic_dawn}

After the star formation begins, their radiations start to heat and ionize the IGM. The Ly$\alpha$ photons from the stars can cause the hyperfine transition in the ground state of the neutral hydrogen, known as Wouthuysen-Field coupling \cite{1952AJ.....57R..31W, 1959ApJ...129..536F}, resulting in the coupling between spin temperature and IGM gas temperature. Therefore, when $T_s<T_R$, we expect an absorption signal at redshifts $z\lesssim 30$ till the universe becomes ionized again \cite{Furlanetto:2006jb, Furlanetto:2006tf, Pritchard:2011xb}.

The Ly$\alpha$ coupling coefficient $(x_{\alpha})$ depends on the star formation history. For a detailed review, follow Refs. \cite{Furlanetto:2006jb, Furlanetto:2006tf, Pritchard:2011xb}. In this work, we consider a simplistic modeling where $x_{\alpha}$ is parameterized using $\tanh$ parameterization \cite{Harker:2015uma, Harker:2011et, Mirocha:2015jra, PhysRevD.98.103529, PhysRevD.110.123506}. Authors in Ref. \cite{Harker:2015uma, Harker:2011et, Mirocha:2015jra}, have used a Markov Chain Monte Carlo (MCMC) technique to extract the global 21-cm signal in the presence of foreground and used successive $\tanh$ parameterization to model the Ly$\alpha$ coupling and X-ray heating of the IGM. The $tanh$ approach can mimic the shape of global 21-cm signals extremely well and can be immediately related to the physical properties of the IGM \cite{Mirocha:2015jra}.
The $\tanh$ parameterization can be expressed as \cite{PhysRevD.98.103529}

\begin{equation}
    \mathcal{L}_i = \mathcal{L}_{(i,\rm ref)}\left(1+ \tanh \left[\frac{z_{i} - z}{\delta z_{i}}\right]\right),
    \label{eq:tanh}
\end{equation}

where, $\mathcal{L}_{(i,\rm ref)}$, $z_{i}$, and $\delta z_{i}$ represent the amplitude, pivot redshift, and duration, respectively. Following Ref. \cite{PhysRevD.98.103529}, we define the Ly$\alpha$ coupling as $x_{\alpha} = 2\mathcal{L}_{\alpha}/(1+z)$ and considered the fiducial values of $\{\mathcal{L}_{(\alpha,\, \rm ref)}, z_{\alpha}, \delta z_{\alpha}\}$ as $\{100, 17, 2\}$. In the next section, we discuss the thermal and ionization evolution of the IGM in the presence of energy deposited from the decaying SCS and X-ray heating.

\section{Evolution of gas in the presence of cosmic strings} \label{sec: gas evolution}

The thermal evolution of the IGM in the absence of energy injection from any exotic source is given by \cite{Peebles:1968ja, Seager:1999bc, Seager:1999km},

\begin{alignat}{2}
    \label{T_b}
    \frac{dT_{gas}}{dz} = \frac{2~T_{gas}}{(1+z)}-\frac{\Gamma_C}{(1+z)H(z)}(T_{\gamma}-T_{gas}),
\end{alignat}
where $H(z)$ represents the Hubble parameter. Here $N_b^{tot}=n_H(1+x_{He}+x_e)$ is the total baryon number density. The ionization fraction is defined as ${n_e}/{n_H}$, and $x_{He}={n_{He}}/{n_H}$ is the helium fraction, where $n_e$, $n_H$, and $n_{He}$ are electron, hydrogen, and helium number density, respectively. Further, the Compton scattering rate $\Gamma_C$ is given by
\begin{equation*}
    \Gamma_C= \frac{8\sigma_Ta_rT_\gamma^4x_e}{3m_e(1+x_{He}+x_e)},
\end{equation*}
where $\sigma_T=6.65\times10^{-25}$ cm$^2$ is the Thompson scattering cross section and the radiation constant $a_r=7.5657\times10^{-16}$J m$^{-1}$K$^{-4}$. The evolution of the ionization fraction is given by \cite{Peebles:1968ja, Seager:1999bc, Seager:1999km}

\begin{alignat}{2}
    \label{Xe}
       \frac{dx_e}{dz} = \frac{\cal C}{(1+z)H(z)}\left[n_HA_Bx_e^2 - 4(1 - x_e)B_B e^{{-3E_{0}}/{4T_{\gamma}}}\right]\, ,
\end{alignat} 
where $E_{0}= 13.6\,$eV is the ground state energy and $E_\alpha\approx {3}/{4}\,E_0$ is the energy of $\rm Ly\alpha$ photon \cite{Ali-Haimoud:2010hou, Ali-Haimoud:2010tlj}. $\cal C$ is the Peebles coefficient, which can be expressed as $\frac{{3}/{4}\,R_{Ly\alpha} + {1}/{4}\, \Lambda_{2s,1s}}{B_B + {3}/{4}\, R_{Ly\alpha} + {1}/{4}\, \Lambda_{2s,1s}}$ \cite{Peebles:1968ja, DAmico:2018sxd}. Here, $ R_{Ly\alpha}=\frac{8\pi H}{3n_H(1-x_e){\lambda_{Ly\alpha}^3}}$ represents the escape rate of Ly$\alpha$ photons, while $\lambda_{Ly\alpha}$ is the Lyman-$\alpha$ wavelength. $\Lambda_{2s,1s}=8.22 \,\mathrm{sec^{-1}}$ is the hydrogen two-photon decay rate. $B_B(T_\gamma)$ is the case-B photo-ionization rate, given by \cite{Peebles:1968ja, Seager:1999bc, Seager:1999km, Bhatt:2019lwt}
 \begin{equation*}
     B_B= A_B\frac{2\pi\mu_eT_\gamma}{4h^3}e^{E_1/T_\gamma}\, ,
 \end{equation*}
where, $E_1=3.4\,$eV is the ionization energy of the first excited state of a hydrogen atom, and $A_B(T_{gas})$ is the case-B recombination rate, which can be expressed as \cite{Peebles:1968ja, Seager:1999bc, Seager:1999km, Bhatt:2019lwt},
 \begin{equation*}
     A_B=\frac{at^b}{1+ct^d}10^{-19}\mathrm{m^3 sec^{-1}}\, .
 \end{equation*}
Here, $t={T_{gas}}/{10^4\,\rm K}$, $a=4.309$, $b=-0.6166$, $c=0.6703$, and $d=0.53$.

We then incorporate the heating of the IGM from X-ray radiation into Eqs. \eqref{T_b} and \eqref{Xe}. In Sec. \eqref{sec: cosmic_dawn}, we described the formulation of Ly$\alpha$ coupling using a $\tanh$ prescription. Similarly, we adopt a $\tanh$ parameterization for X-ray heating of the IGM temperature and ionization fraction. Following Ref. \cite{PhysRevD.98.103529}, we define $\mathcal{L}_{xe}$ and $\mathcal{L}_X$ as the contributions to the ionization fraction and IGM temperature, respectively, due to X-ray heating. Here, $\mathcal{L}_{xe}$ and $\mathcal{L}_X$ are formulated analogously to Eq. \eqref{eq:tanh}. Now, the modified form of Eqs. \eqref{T_b} and \eqref{Xe} can be expressed as 
\begin{alignat}{2}
    \frac{dT_{gas}}{dz} & = \frac{dT_{gas}}{dz}\Bigg{|}_{\rm Eq. \eqref{T_b}} + \frac{d\mathcal {L}_X}{dz}\, , \label{eq:T_b_xray} \\
    \frac{dx_e}{dz} & = \frac{dx_e}{dz}\Bigg{|}_{\rm Eq. \eqref{Xe}}+\mathcal {L}_{xe}\, . \label{eq:x_e_xray}
\end{alignat}

The free parameters and their fiducial values associated with $\mathcal L_{xe}$ are $\{\mathcal L_{(xe,\,\rm ref)}, z_{xe}\,\text{and}\,\delta z_{xe}\}$ and $\{1, 9, 3\}$, respectively. Similarly, for $\mathcal L_{X}$ the free parameters and their fiducial values are $\{\mathcal L_{(X,\rm ref)}, z_{X}\,\text{and}\,\delta z_{X}\}$ and $\{1000\,\textrm{K}, 12.75, 1\}$, respectively \cite{PhysRevD.98.103529}.

The thermal and ionization fraction evolution of the IGM can be modified in the presence of energy injection from exotic sources. Earlier in Sec. \eqref{sec.SCS}, we discussed how decaying SCS can emit copious electromagnetic radiations that can potentially ionize and heat the IGM. In the presence of these radiations, the IGM temperature $(T_{gas})$ and ionization fraction $(x_e)$ can increase significantly. The evolution of the IGM temperature in the presence of any exotic source of energy injection \cite{Ali-Haimoud:2010hou, Mu_oz_2015, Mitridate:2018iag, Bhatt:2019qbq, 10.1093/mnras/stad2457, Natwariya:2021PMFAE, Natwariya:2022xlv, Natwariya:2024ML, PhysRevD.111.043002}

\begin{align}
    \frac{dT_{\text{gas}}}{dz} 
    &= \frac{dT_{\text{gas}}}{dz}\Bigg|_{\rm Eq.~\eqref{eq:T_b_xray}} 
    - \frac{2}{3(1+z)H(z)} \frac{(1 + 2x_e)}{3N_b^{\text{tot}}} \notag \\
    &\qquad \qquad \qquad \qquad \qquad \qquad \qquad \times \frac{d^2E}{dV\,dt}\Bigg|_{\text{dep}} \,, \label{eq:T_b_modified} \\
    \frac{dx_e}{dz} 
    &= \frac{dx_e}{dz}\Bigg|_{\rm Eq.~\eqref{eq:x_e_xray}} 
    - \frac{1 - x_e}{(1+z)H(z)N_b^{\text{tot}}} \notag \\
    &\qquad \qquad \qquad \times \left( \frac{\mathcal{C}}{E_0} + \frac{1 - \mathcal{C}}{E_\alpha} \right) 
    \frac{d^2E}{dV\,dt}\Bigg|_{\text{dep}} \,. \label{eq:x_e_modified}
\end{align}

The energy density deposition rate, $d^2E/dVdt$, is taken from Eq. \eqref{eq: energy deposition}. We follow the ``SSCK" approximation, in which $(1-x_e)/3$ fraction of energy ionizes, while $(1+2x_e)/3$ fraction of energy heats the IGM \cite{1985ApJ...298..268S, PhysRevD.70.043502}. In the next section, we solve the modified thermal and ionization evolution equations simultaneously to investigate the effect of decaying SCS on the global 21-cm signal.

\begin{figure*}
    \begin{center}
        \subfloat[] {\includegraphics[width=3.5in,height=2.5in]{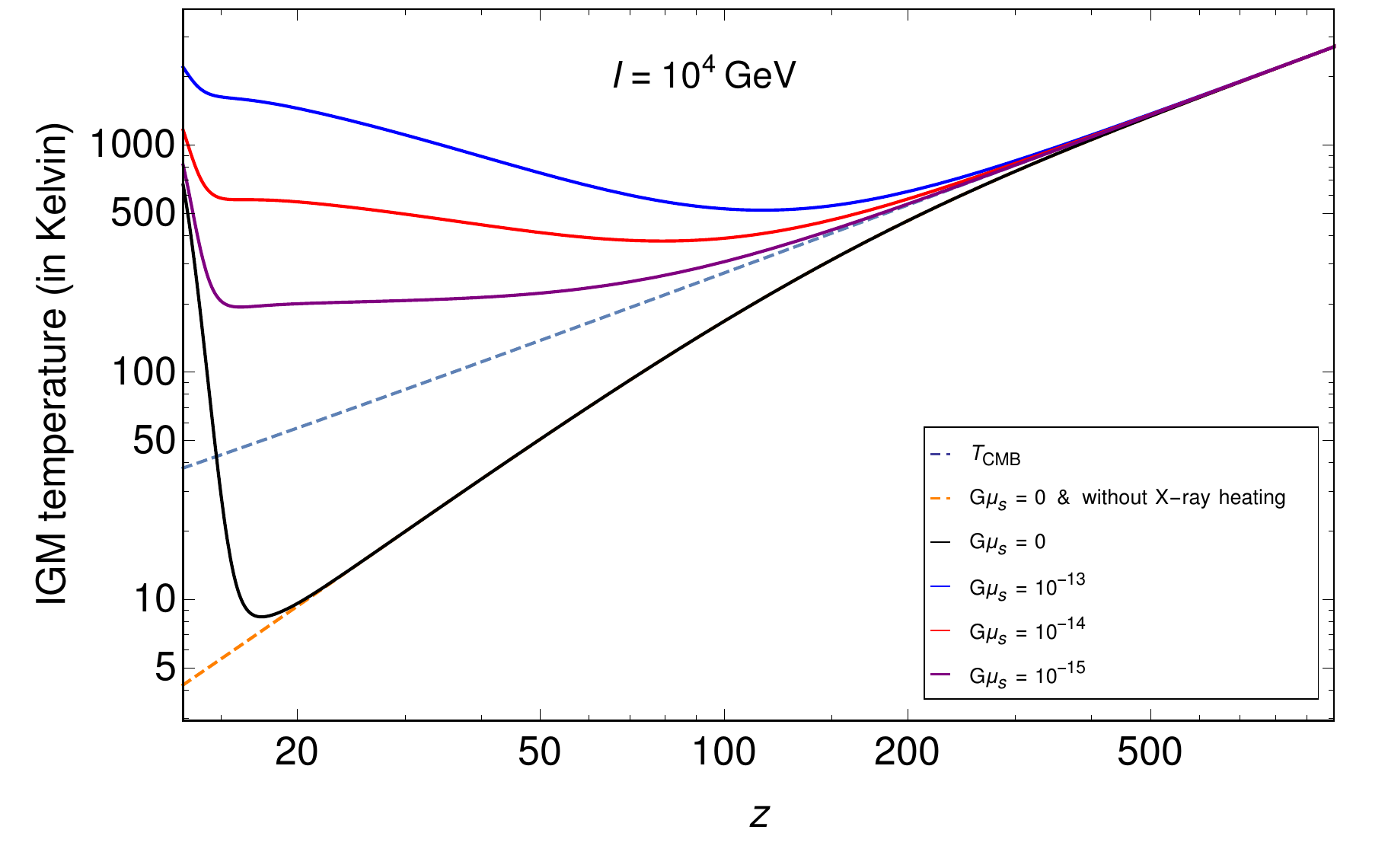}\label{p2a}}
        \subfloat[] {\includegraphics[width=3.5in,height=2.5in]{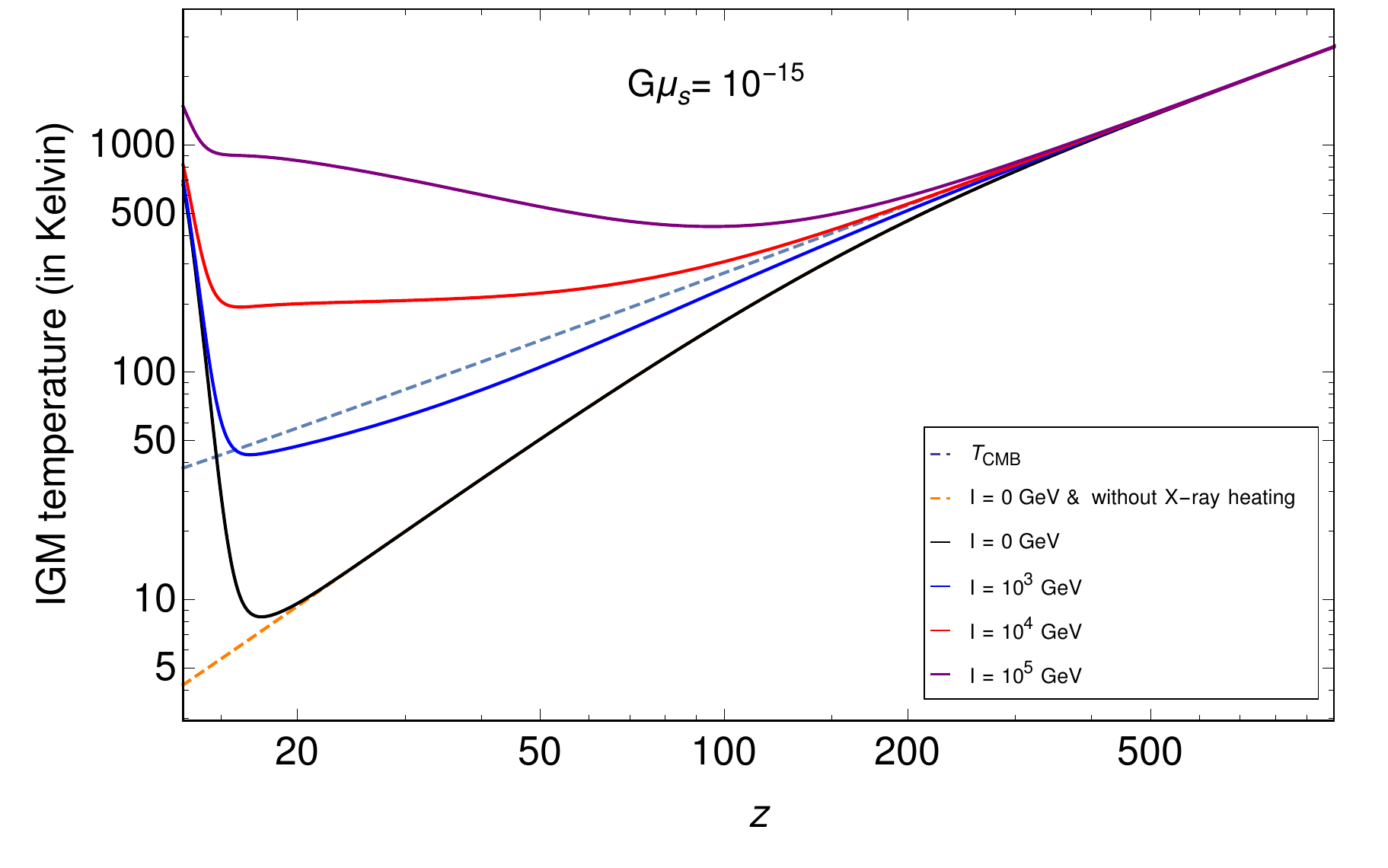}\label{p2b}}
        
    \end{center}
    \caption{Represents the effect of decaying superconducting cosmic string on IGM temperature ($T_{gas}$) with redshift ($z$). The grey and orange dashed lines depict the CMB temperature and IGM gas temperature evolution without X-ray heating in the absence of SCS radiation, while the solid black line represents the IGM gas temperature evolution with X-ray heating in the absence of SCS radiation. In the left panel [Fig. (\ref{p2a})], we vary the dimensionless string tension for a fixed string loop current $I=10^{4} $ GeV. In the right panel [Fig. (\ref{p2b})], we fixed $G\mu_s=10^{-15}$ and vary the cosmic string current.}
    \label{plot: Tb_plot}
\end{figure*}

\begin{figure*}
\begin{center}
        \subfloat[] {\includegraphics[width=3.5in,height=2.5in]{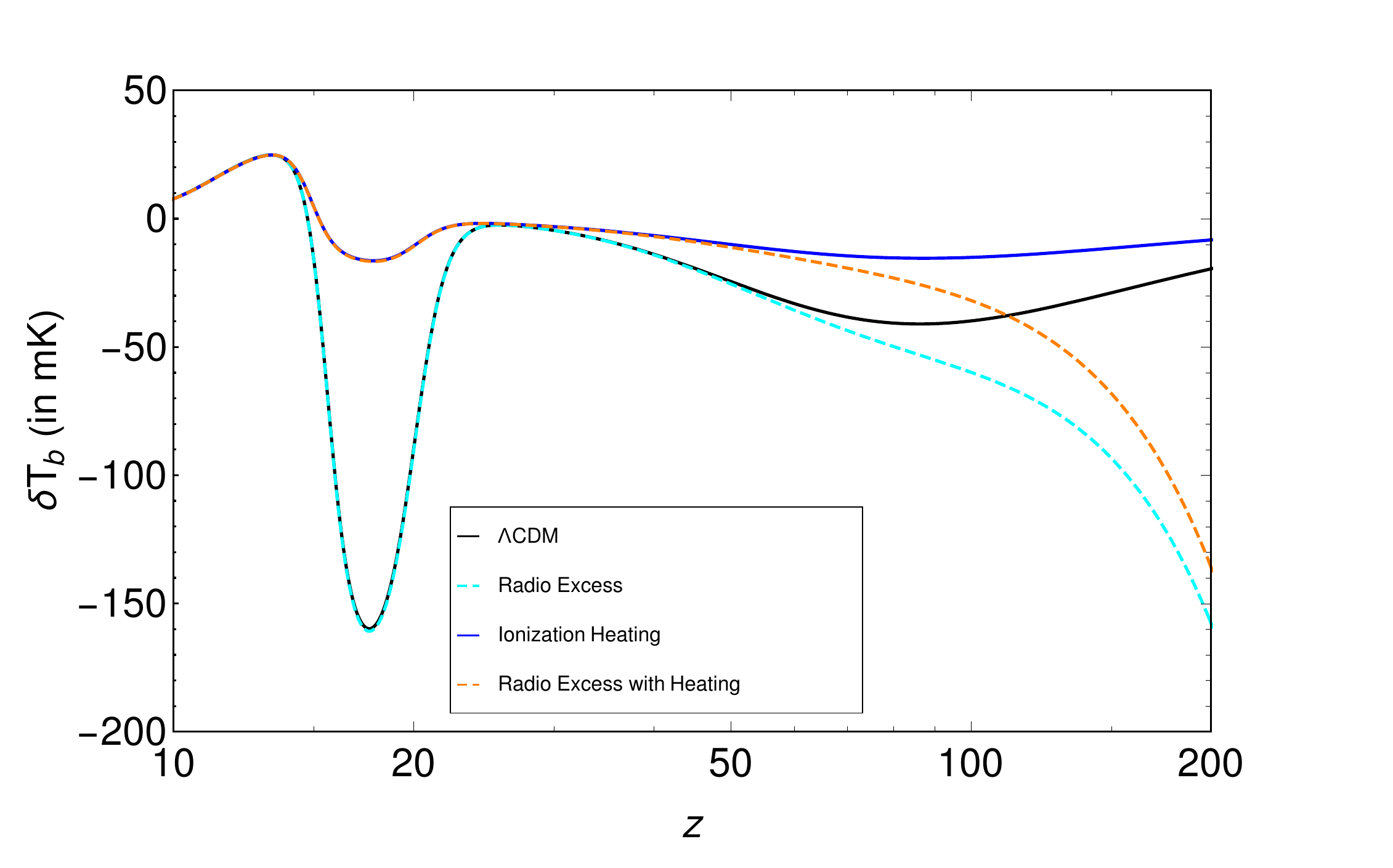}\label{t21a}}
        \subfloat[] {\includegraphics[width=3.5in,height=2.5in]{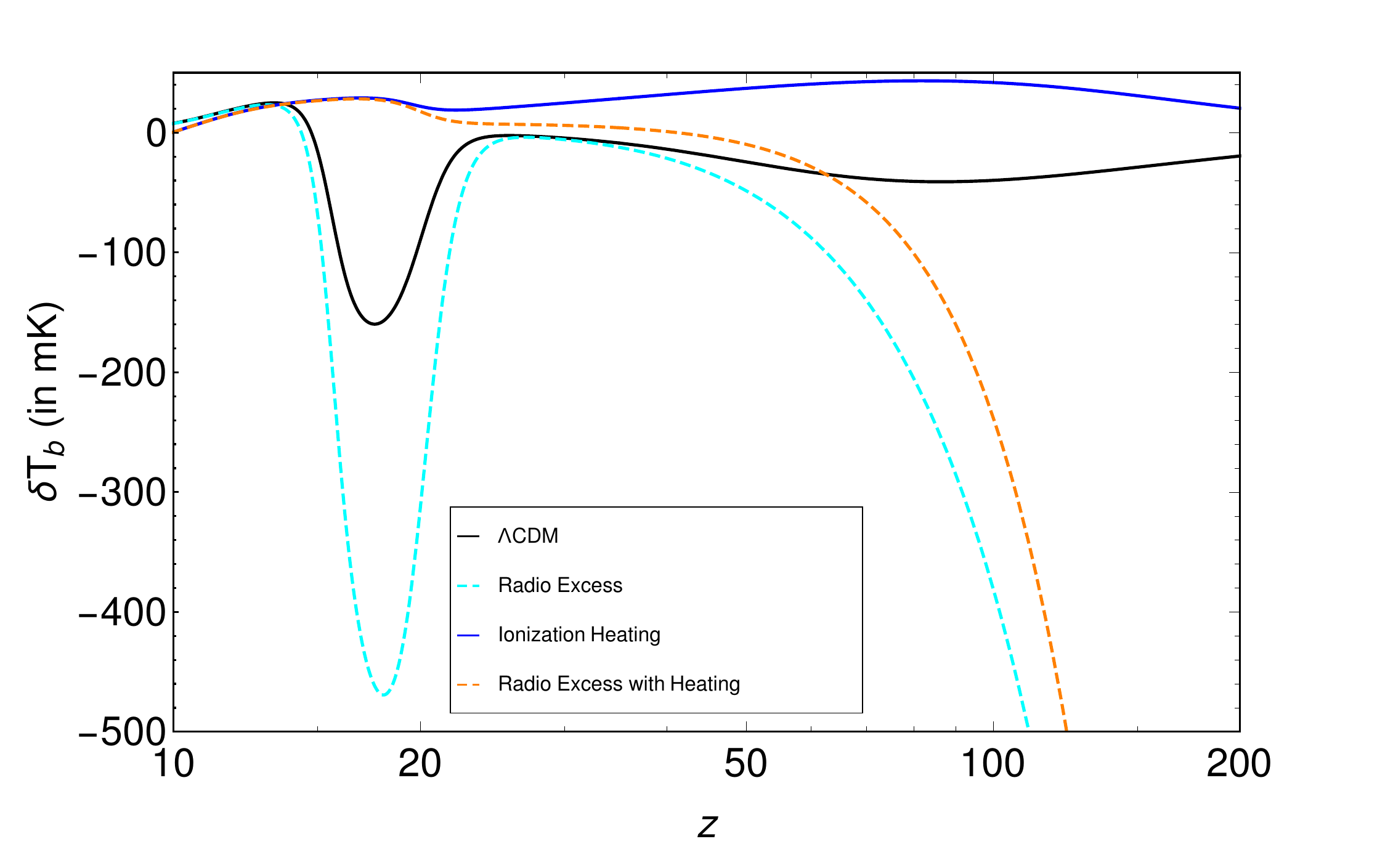}\label{t21b}}
        
    \end{center}
\caption{The evolution of the differential brightness temperature $\delta T_{b}$ in the presence of superconducting cosmic strings. In both the figures, the black solid line represents $\delta T_b$ from the standard $\Lambda\rm CDM$ framework. In Fig. (\ref{t21a}) we consider strings with loop current $I= 10^3$ GeV and $G\mu_s = 5\times 10^{-16}$ and in Fig. (\ref{t21b}) we consider strings with loop current $I= 10^6$ GeV and $G\mu_s = 5\times 10^{-16}$. The cyan-dashed and blue solid lines represent $\delta T_b$ when we separately consider radio photons (due to first term on the RHS of Eq. \eqref{eq: cs_tot_energy}) and ionizing photons from SCS radiation (due to third term on the RHS of Eq. \eqref{eq: cs_tot_energy}), respectively. The orange dashed line represents $\delta T_b$ evolution on considering the entire photon spectrum emitted from the strings simultaneously (due to the first and third term on the RHS of Eq. \eqref{eq: cs_tot_energy})).}
\label{plot: t21}
\end{figure*}

\section{Results} \label{sec: results}

In this section, we investigate the impact of the decaying SCS on the global 21-cm absorption signal and obtain constraints on $I$ and $G\mu_s$. We study the thermal evolution of IGM in the presence of decaying SCS ($d^2E/dVdt = 0$) by solving Eqs. \eqref{T_b} and \eqref{Xe} simultaneously with the initial conditions $T_{gas} = 2758\text{ K}$, and $x_e = 0.05725$ at redshift $z = 1010$ adopted from \texttt{Recfast++} \cite{10.1111/j.1365-2966.2010.16940.x, 10.1111/j.1365-2966.2010.17940.x}. 
The evolution of IGM temperature with redshift $z$ in the presence of SCS radiation with the X-ray heating is shown in Fig.(\ref{plot: Tb_plot}). The gray-dashed line represents the CMB temperature, the orange dashed line and the solid black line indicate the evolution of the IGM gas temperature in the absence of cosmic strings. At redshifts $30\lesssim z\lesssim 200$, the $T_{gas}$ evolves adiabatically after decoupling from the CMB. The rise in $T_{gas}$ at redshifts $z\lesssim 20$ indicates heating of the IGM due to X-ray radiation (Eq. \ref{eq:T_b_modified}). We then include the energy injection from the decaying SCS. We consider $\mathcal {F}(z,\omega)$ shown in Eq. \eqref{eq: energy deposition} to be unity, suggesting an instantaneous deposition of energy \cite{PhysRevLett.121.011103, PhysRevD.93.023521}. In Fig. (\ref{p2a}), we plot $T_{gas}$ for different values of dimensionless string tension $G\mu_s= 10^{-13}, 10^{-14}$ and $10^{-15}$ while keeping the string loop current $I=10^4$ GeV fixed-- shown in the blue, red, and purple solid lines, respectively. We find that, on increasing the string tension, $T_{gas}$ increases significantly. For certain values of $G\mu_s$ and $I$, for instance $G\mu_s =10^{-15}$ and $I = 10^4\,\rm GeV$, the IGM temperature can even rise above CMB temperature. This can be observed by analyzing Eqs. \eqref{Gamma2} and \eqref{eq: energy deposition}. It can be seen that $d^2E/dVdt$ is directly proportional to $\omega^{1/3}$ and $(G\mu_s)^{7/12}$ for $I>I^*$, while $\propto t^{-19/6}$ which can translate to $\propto (1+z)^{19/4}$ in the matter-dominated era. However, the number density rate of photons with energy $\omega$ falls as $\propto \omega^{-5/3}$ (Eq. \ref{eq: cs_photon_spectrum}). Further, in Fig. (\ref{p2b}), we fix $G\mu_s= 10^{-15}$ and vary the string loop current $I= 10^{3}, 10^{4}$ and $10^{5}$ GeV-- depicted in the blue, red, and purple solid lines, respectively. In the matter-dominated era, the energy deposition rate ${d^2E}/{dVdt}\propto I^{5/6}$ for $I>I^*$ (see Eqs. \ref{Gamma2} and \ref{eq: energy deposition}). As a result, the energy injection rate increases for large $I$ values. Next, we study the effect of superconducting cosmic strings on the global 21-cm signal.

\begin{figure}[htbp]
        \centering
        \includegraphics[width=\linewidth]{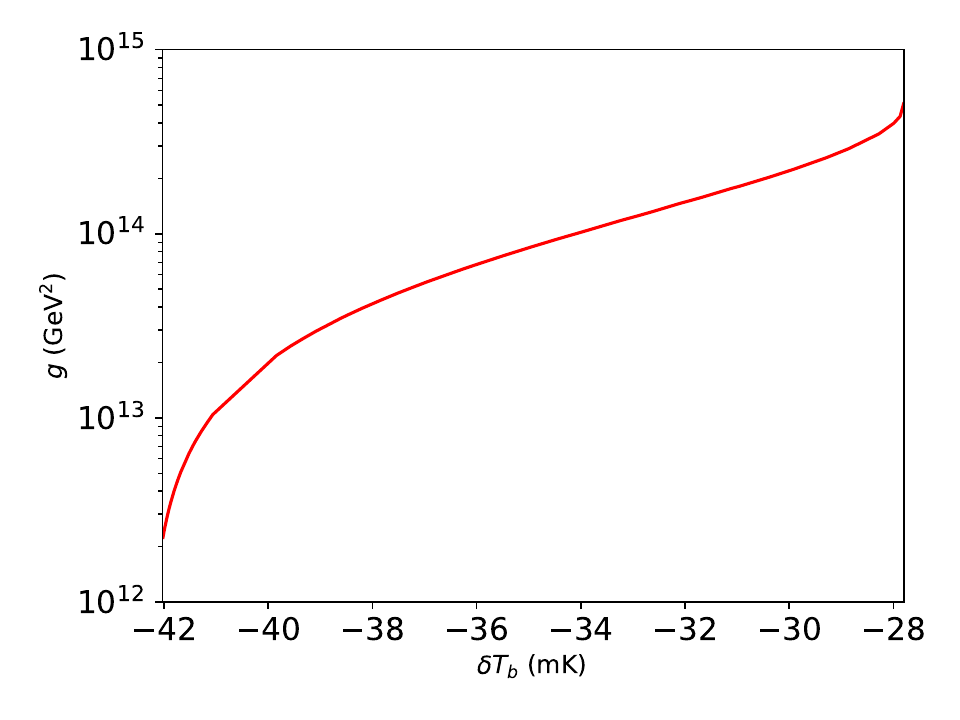}
        \caption{Constraints on the parameter $g$ as a function of the 21-cm differential brightness temperature $\delta T_b$ at $z\sim89$. Here the parameter $g= I^2 \big(\Gamma~G\mu_s\big)^{-7/6}$. The parameter value above $g\gtrsim5.1\times10^{14}\, \rm GeV^2$ is excluded due to emission signal at cosmic dawn, and below $g\lesssim2.2\times10^{12}\, \rm GeV^2$, the effect of decaying SCS on 21-cm signal is no longer detectable.} 
        \label{fig:g}
    \end{figure}

\begin{figure}
\centering
\includegraphics[width=\linewidth]{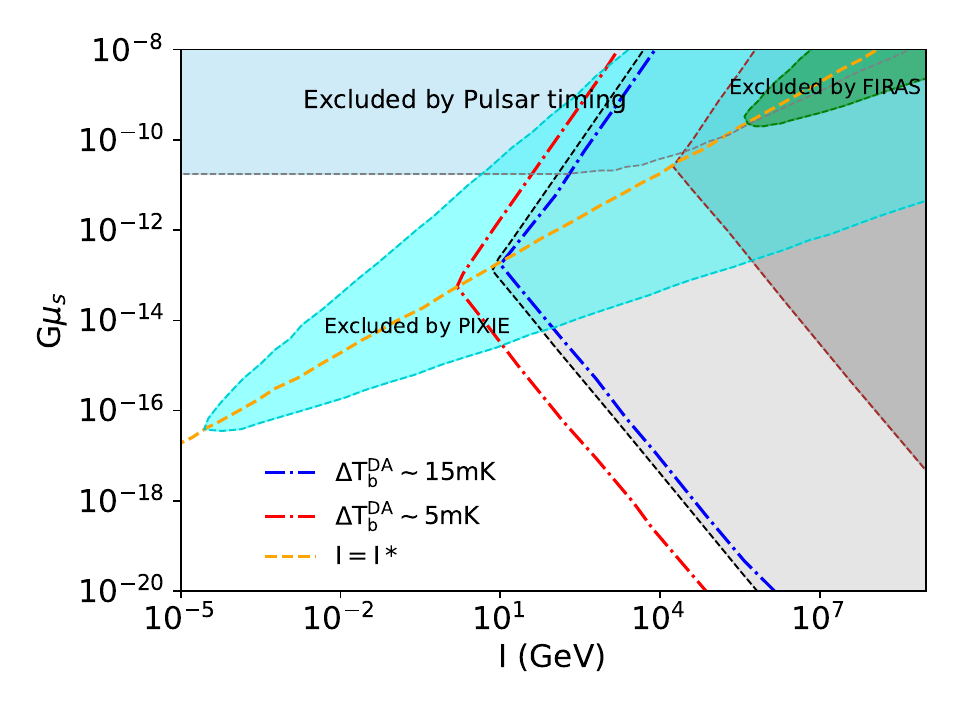}
\caption{Constraints on the cosmic string parameter space from the 21-cm signal of the dark ages. The dashed orange line indicates the critical current corresponding to $G\mu_s$.
The black and brown dashed lines are the constraints from Gessey-Jones et al. (2024) at $1\sigma$ and $2\sigma$, respectively \cite{10.1093/mnras/stae512}. The grey-shaded regions indicate the excluded region by Gessey-Jones et al. (2024) \cite{10.1093/mnras/stae512}. The sky-blue shaded region indicates the excluded region by the pulsar timing constraint from gravitational radiation obtained by Ref. \cite{Miyamoto:2012ck}. The cyan and green shaded regions were obtained by the COBE/FIRAS and PIXIE measurements at the $2\sigma$ limit \cite{Cyr:2023iwu}. }
\label{plot: contour}
\end{figure}

In Fig. (\ref{plot: t21}), we have shown the impact of decaying SCS on the evolution of the global 21-cm signal (see Eq. \ref{T21}). In both figures, the solid black lines show the evolution of $\delta T_b$ in a $\Lambda\rm CDM$ framework without cosmic strings. The $\delta T_b$ takes value of $\sim -42\,\rm mK$ and $\sim -160\,\rm mK$ at redshifts $z = 89$ and $z = 17$, respectively, for the fiducial values considered in the $tanh$ parameterization (Eq. \ref{eq:tanh}), in the $\Lambda\rm CDM$ framework. The amplitude of $\delta T_b$ during the cosmic dawn era $(z \sim 17)$ can vary for different values of the free parameters associated with the Ly$\alpha$ coupling and X-ray, which would indicate different star formation scenarios. However, as this work focuses on the heating of IGM and CMB from decaying superconducting cosmic strings, we fixed those fiducial values. In Fig. \eqref{t21a}, we consider SCS with loop current $I=10^3$ GeV and string tension $G\mu_s=5\times 10^{-16}$.
First, we consider only the nonthermal radio photons produced from these strings and find an increase in the background radio radiation $(T_{R})$ at redshifts $z\gtrsim 50$--- shown in the cyan dashed line. This can be analyzed from Eq. \eqref{rho21} and Eq. \eqref{t21}, where the redshift dependence of $T_{21}^{\rm{SCS}}$ (temperature of nonthermal photons produced from decaying SCS) follows $\propto(1+z)^{13/4}$. Therefore, $T_{21}^{\rm{SCS}}$ is greater during the dark ages era $(z\sim 89)$ compared to the cosmic dawn $(z\sim 17)$. We then consider only the ionizing radiation while fixing the background radiation to CMB, $T_R=T_\gamma$. This results in an increase in the IGM temperature $(T_{gas})$ that leads to a shallower $\delta T_b$ at redshifts $z\sim 17$ and $z\sim 89$--- shown in the blue solid line. Finally, we considered both the nonthermal radio and ionizing photons produced from these strings and plotted $\delta T_b$--- shown in the orange dashed line. In Fig. \eqref{t21b}, we considered a larger loop current $(I= 10^6\,\rm GeV)$ for the same string tension and present the variations in $\delta T_b$. We first consider only the nonthermal radio photons and find an enhanced absorption signal at $z\sim 17$ and $z\sim 89$--- shown in the cyan solid line. Then, we consider only the ionizing radiations and find that such SCS can potentially erase the cosmic dawn and dark ages 21-cm signal--- shown in the blue solid line. Lastly, we consider both radiation spectra together and find that such strings can produce a large absorption signal during the dark ages; however, they can potentially erase the cosmic dawn signal--- shown in the orange dashed line. This has primarily been ignored or lacks explicit consideration in the previous studies \cite{Brandenberger:2019lfm, Theriault:2021mrq, 10.1093/mnras/stad2457, Acharya:2023ygd, 10.1093/mnras/stae512}. We restrict such scenarios as the emission or erasing of the cosmic dawn signal would imply an early reionization of the universe. In the next section, we limit the SCS parameter space by analyzing the dark ages signal.


 In Fig. (\ref{fig:g}), we derived the bound on the decay efficiency parameter $g(I, G\mu_s)$ as a function of the 21-cm differential brightness temperature $\delta T_b$. At redshift $z \sim 89$, the standard $\Lambda\rm CDM$ model predicts $\delta T_b \approx -42\, \rm mK$. Injected energy from decaying SCS can increases the $\delta T_b$, and at $\delta T_b\sim-27.8\, \rm mK$ the parameter reach at $g\sim5.1\times10^{14}\, \rm GeV^2$. The values above $g\gtrsim5.1\times10^{14}\, \rm GeV^2$ are excluded due to the presence of an emission signal at cosmic dawn ($z\sim17$), and below $g\lesssim2.2\times10^{12}\, \rm GeV^2$, the effect of decaying SCS on the 21-cm signal is no longer detectable. For a fixed $\delta T_b$, the corresponding energy injection is constant; consequently, the parameter $g(I, G\mu_s)$ becomes a constant. For a constant $g(I, G\mu_s)$ value, one can obtain bounds for the string current by fixing the string tension, or vice versa, using Eqs. \eqref{eq: gmu_I<I*} and \eqref{eq: gmu_I>I*}. In this case, the degeneracy between the string current $I$ and tension $G\mu_s$ then appears as a straight line on a log-log plot in the ($I$, $G\mu_s$) plane. The slope of the straight line varies for two different regimes, $I<I^*$ and $I>I^*$, due to the different functional form of $\Gamma$ in Eq. \eqref{Gamma2}. The transitions between these two regimes occur along the curve defined by $I=I^*(G\mu_s)$.

In Fig. \eqref{plot: contour}, we derive upper bounds on $G\mu_s$ and $I$ from the dark ages 21-cm signal. In Sec. \eqref{sec:level1}, we have explained that for an observational integration time of 20,000 and $10^5$ hours, the uncertainty in the detection of the standard dark ages $\delta T_b$ signal becomes 15 mK and 5 mK, respectively, for future lunar-based experiments \cite{Burns:2020gfh, Rapetti:2019lmf}. Therefore, to constrain $I$ and $G\mu_s$, we take the amplitude to be $\delta T_b = -36$ mK and $-26$ mK at $z\sim 89$, such that the change in $\delta T_b$ $(\Delta T_b)$ due to decaying SCS will become $5\,\rm mK$ and $15\,\rm mK$, respectively. The blue dash-dotted line shows the upper bounds on $ G\mu_s$ and $I$ for $\Delta T_b\sim 15\, \rm mK$, whereas the red dash-dotted line represents $\Delta T_b = 5$ mK. We find that even after considering the heating of the IGM due to decaying SCS, the dark ages signal can provide stronger and astrophysical uncertainty-free upper bounds on cosmic strings. For example, considering the $(\Delta T_b)$ to $5$ mK, varying the cosmic string tension from $1\times10^{-20}$ to $\sim5.4\times10^{-14}$ the upper bound on cosmic strings loop current varies from $\sim7.3\times10^4$~GeV to $\sim1.5$~GeV. Further increasing the string tension from $\sim5.4\times10^{-14}$ to $1\times10^{-8}$, the upper bounds on the loop current get relaxed and change from $\sim1.5$~GeV to $\sim1.6\times10^3$~GeV.

 To further compare our results with previously excluded regions, we have shown the constraints from T. Gessey-Jones et al. (2024) \cite{10.1093/mnras/stae512} in Fig. (\ref{plot: contour}). The authors jointly consider the upper bound on the 21-cm power spectrum from HERA, the global 21-cm signal from SARAS3 during the cosmic dawn and the epoch of reionization. In this study, they consider only the contributions to the excess radio background produced from decaying SCS. Furthermore, the authors performed a Bayesian analysis where they marginalized over astrophysical uncertainties such as star formation efficiency and X-ray heating to find an upper bound on the SCS properties. The grey-shaded region with black and brown dashed lines shows the upper bound with 68\% $(1\sigma)$ and 95\% $(2\sigma)$ confidence level, respectively \cite{10.1093/mnras/stae512}. The sky-blue shaded region depicts the excluded region from the pulsar timing array (PTA) on measuring gravitational radiation in Ref. \cite{Miyamoto:2012ck}. In this study, the authors consider the stochastic gravitational wave background (SGWB) produced by the SCS loops. They simulate the gravitational wave emission from SCS loops and compare the predicted SGWB with the upper limits set by PTA observations. The cyan and green shaded regions are obtained from the Cosmic Background Explorer / Far Infrared Absolute Spectrophotometer (COBE/FIRAS) and forecasted Primordial Inflation Explorer (PIXIE) measurements at the $2\sigma$ limit \cite{Cyr:2023iwu}. In this study, the authors constrain the SCS parameter space using FIRAS and PIXIE bounds on $\mu_s$ types of CMB spectral distortion due to energy injection into the IGM from the decaying SCS. 
If the energy injected into the IGM at the redshift between $5\times10^4\lesssim z\lesssim2\times10^6$ results in $\mu_s$-type of CMB spectral distortions, and when the energy injected at redshifts below $5\times10^4$ results in $y$-type spectral distortions in CMB. The COBE/FIRAS put the upper bounds on the $\mu_s$-type ($y$-type) distortion to be $\mu = 9.0 \times10^{-5}$ $(y = 1.5\times10^{-5})$ \cite{Fixsen:1996nj, Natwariya:2025ftu}. The future experiment PIXIE aims to detect $\mu_s$-type ($y$-type) distortion at levels of $\mu=5\times10^{-8}$ ($y=10^{-8}$) at a $5\sigma$ level \cite{Kogut:2011xw, Natwariya:2025ftu}. Additionally, the orange dashed line shows that the critical current varies with the $G\mu_s$ value (see Eq. \ref{I*}). Below this line, power emitted as electromagnetic radiation dominates, whereas gravitational radiation is more important above the line.

\section{Discussion and conclusion} \label{sec: conclusions}

Decaying superconducting cosmic strings (SCS) can emit both ionizing and radio photons after the recombination.
These ionizing photons can alter the thermal and ionization evolution of the IGM. Conversely, nonthermal radio photons from SCS can increase the background radiation temperature.
In this study, we investigate the influence of the decaying SCS on the global 21-cm signal during the dark ages. 
The dark ages global 21-cm signal is independent of astrophysical uncertainties, making it an ideal probe for any exotic physics after recombination. 
Future proposed space and lunar-based experiments such as FARSIDE \cite{burns2019farsidelowradiofrequency}, DAPPER \cite{Burns:2021ndk}, LuSee Night \cite{bale2023luseenightlunarsurface}, and SEAMS \cite{borade2021fpga} may measure this signal.
The recent proposal for LuSee Night to reach the far side of the moon in 2026 aims to observe the sky in the frequency range of $0.1-50\,\rm MHz$, which may allow for the detection of the global 21-cm signal from the dark ages \cite{10906958}. 
Moreover, for future lunar-based experiments, an integration time of 20,000 hours is anticipated to achieve an uncertainty (\(\Delta T_b\)) of 15 mK in detecting the standard dark ages signal. Furthermore, extending the integration time to 100,000 hours can reduce the uncertainty to 5 mK \cite{Burns:2020gfh, Rapetti:2019lmf}.

We present upper bounds on the SCS parameter space in Fig. \eqref{plot: contour} by considering that SCS can alter the amplitude of the global 21-cm signal $(\Delta T_b)$ by $5$ mK and $15$ mK. For example, considering the $(\Delta T_b)$ to $5$ mK, varying the cosmic string tension from $1\times10^{-20}$ to $\sim5.4\times10^{-14}$ the upper bound on cosmic strings loop current varies from $\sim7.3\times10^4$~GeV to $\sim1.5$~GeV. Further increasing the string tension from $\sim5.4\times10^{-14}$ to $1\times10^{-8}$, the upper bounds on the loop current get relaxed and change from $\sim1.5$~GeV to $\sim1.6\times10^3$~GeV. We have also presented the available constraints on the SCS parameter space for comparison.

Before concluding our work, we specify that to derive upper bounds on the $I$ (string current) and G$\mu_s$ (string tension), we adopted a fiducial cosmological model with Planck 2018 best-fit parameters. Further, we considered that if the difference between the amplitude of the global 21-cm signal from the adopted fiducial cosmological model and in the presence of superconducting cosmic strings (SCS) is greater than the instrumental noise level, then that could hint towards the presence of SCS. However, as the dark ages do not host luminous astrophysical objects, variations in the dark ages signal due to cosmological parameter uncertainties become important. In Ref. \cite{Okamatsu:2023diy, Mohapatra:2024djd}, the authors have shown the variation in the dark ages $\delta T_{b}$ amplitude. In Ref. \cite{Mohapatra:2024djd}, the authors have shown that on varying the cosmological parameters $(\Omega_bh^2,\Omega_mh^2,Y_p)$ within their $1\sigma$ values, the $\delta T_{b}$ amplitude can vary by $\Delta T_{b} \sim 6~\rm mK$ $(\delta T_{b} \simeq -42_{-2}^{+4}~\rm mK)$. Notably, the $\delta T_{b}$ amplitude can take a value of $\sim -38~\rm mK$ for certain values of $\Omega_bh^2,\Omega_mh^2$ and, $Y_p$. The energy injection from decaying SCS will reduce the $\delta T_{b}$ amplitude $(\delta T_{b}\lesssim -42~\rm mK)$--- as shown in Fig. \ref{t21a}. As we discussed earlier that the future lunar-based experiments with observation time of $20,000$ and $10^5$ hours, may achieve an uncertainty of $(\Delta T_b\simeq)$ 15 mK and 5 mK, respectively, in detecting the standard dark ages signal. We can interpret that obtaining upper bounds on the $I$ and G$\mu_s$ by $\Delta T_b \sim 5~\mathrm {mK}~(\delta T_b\sim -42_{-5}^{+5}~\rm mK)$ provides conservative bounds. However, a full statistical treatment has not been considered in this work, which would involve finding posterior distributions of SCS and cosmological parameters with respect to the dark ages $T_{21}$ signal by deploying Markov Chain Monte Carlo (MCMC) or Fisher forecast.

Similarly, we also consider an optimistic foreground removal case, i.e., a complete removal of the foreground radiation. However, in practice, the foreground is formidable, and mitigating it requires the exact knowledge of its spectral form and spectral energy distribution of the contributing sources. Additionally, foreground polarization and instrument systematics such as beam chromaticity can add a layer of complexity in the mitigation (see Ref. \cite{Furlanetto:2006jb} for a comprehensive review). Therefore, the conservative bounds on the SCS parameter space derived in this work may vary with marginalization of the cosmological parameter uncertainties and foreground removal pipelines.

\begin{acknowledgments}
The authors would like to thank Sandeep Kumar Acharya and Tripurari Srivastava for the fruitful discussion. This work is in part supported by the Hangzhou City Scientific Research Funding, No. E5BH2B0105/B015F40725006. We also thank the anonymous referees for their valuable suggestions and for improving our manuscript quality.

\end{acknowledgments}

\bibliography{shankar.bib}

\end{document}